\documentclass[12pt]{article}
\usepackage{amssymb,amsmath}
\begin{document}


\sloppy
\title
{\hfill{\normalsize\sf FIAN/TD/01-15}    \\
            \vspace{1cm}
{\Large  On the theory of scalar pair production by a potential barrier }
}
\author
 {
       A.I.Nikishov
          \thanks
             {E-mail: nikishov@lpi.ru}
  \\
               {\small \phantom{uuu}}
  \\
           {\it {\small} I.E.Tamm Department of Theoretical Physics,}
  \\
               {\it {\small} P.N.Lebedev Physical Institute,}
  \\
         {\it {\small} 119991, Leninsky Prospect 53, Moscow, Russia.}
 }
\date{13 Nov. 2001}
\maketitle
\begin{abstract}
The problem of the scalar pair production by a one-dimensional vector-
potential $A_{\mu}(x_3)$ is reduced to the $S-$ matrix formalism of the
theory with an unstable vacuum. Our choice of in- and out-states does not
coincide with that of other authors and we argue extensively in favor of our
choice. In terms of our classification the states that can be created by the
field enter into the field operator in the same way as do the states that
cannot be created by the field, i.e. the field operator has the usual form.
We show that the norm of a solution of  the wave equation is determined by
one of the amplitude of its asymptotic form for $x_3\to \pm\infty$. For the step
potential and for the constant field potential we get the explicit expressions
for the complete in- and out-sets of orthonormalized wave functions. For the
constant electric field we obtain the scalar particle propagator  in terms
of the stationary states and show that with our choice of in- and out-states
it has the form dictated by the general theory.
\end{abstract}

\newpage

\section{Introduction and the choice of in- and out-states}

Pair production by an external field can be treated either in the framework
of $S-$matrix formalism [1-5], or equivalently by the Feynman method using
the propagators [1,6-8]. For the stationary potential the field is not
switched off for $t\to\pm\infty$. So the reduction to the $S-$matrix formalism
requires choosing the in- and out-states. How to do this is briefly shown in [1].
Another choice is made in [9] and accepted in later literature [10,11]. The
correct choice is especially important in dealing with higher order processes,
when the answer is not known in advance from some other considerations. In
this paper we argue extensively in favor of our choice. It is reasonable to
consider the case of scalar particle separately, because the complications
due to the spin are absent here. Besides it is useful to have all the
stages of a more simple case before eyes, when treating the spinor case.

We consider at first the one-dimensional potential $A^0(x_3)$ and assume  for the
beginning that the corresponding electrical field $E_3=-\frac{\partial A^0}
{\partial x_3}$ disappears for $x_3\to\pm\infty$. We use the metric
$$
\eta_{\mu\nu}={\rm diag(-1,1,1,1)}.  \eqno(1)
$$
It is useful to introduce the kinetic energy $\pi^0(x_3)$ and momentum
$\pi_3(x_3)$ of a classical particle defined by the expressions
$$
\pi^0(x_3)=p^0-eA^0(x_3),\quad \pi_3(x_3)=\sqrt{\pi^2_0(x_3)-m^2_{\perp}},
\quad m^2_{\perp}=m^2+p^2_1+p^2_2. \eqno(2)
$$
 The first relation in (2) merely expresses the total energy conservation. We
also use the notation
$$
\left.\pi^0(x_3)\right|_{x_3\to\pm\infty}=\pi^0(\pm), \quad
\left.\pi_3(x_3)\right|_{x_3\to\pm\infty}=\pi_3(\pm)=\sqrt{\pi_0^2(\pm)-
m_{\perp}^2}. \eqno(3)
$$

In contrast to [1] and [5] we assume here that the charge of a scalar particle
$e=-|e|$ in order the analogy  with the electron would be closer. We are interested
here mainly in the states that can be created by the field (Klein region).
Assuming for definiteness $E_3>0$, we have in this region
$$
\pi^0(-)>m_{\perp}^2, \qquad \pi^0(+)<-m_{\perp}^2,\eqno(4)
$$
i.e. large positive $x_3$ are accessible only to antiparticles.

For brevity reasons we write only the wave function factor depending on
$x_3$. Outside the field the particle is free and we first classify the
states by their asymptotic form
$$
\left.{}_{\pm}f_p\right|_{x_3\to-\infty}=
[2\pi_3(-)]^{-\frac12}\exp[\pm i\pi_3(-)x_3],\qquad
\left.{}^{\pm}f_p\right|_{x_3\to\infty}=
[2\pi_3(+)]^{-\frac12}\exp[\pm i\pi_3(+)x_3]. \eqno(5)
$$
The normalization  factors are chosen in such a way that the density current
along the third axis is equal to unity up to a sign. Two sets of functions
in (5) are connected by the relations
$$
{}_+f_p=c'_{1p}\,{}^+f_p+c'_{2p}\,{}^-f_p,
$$
$$
{}_-f_p=c'{}^*_{2p}\,{}^+f_p+c'{}^*_{1p}\,{}^-f_p.  \eqno(6)
$$
The second relation can be obtained from the first one by complex conjugation.
 The current conservation along the third axis gives 
$$
|c'_{1p}|^2-|c'_{2p}|^2=1. \eqno(7)
$$
From (6) and (7) we find the reversed relations
$$
{}^+f_p=c'{}^*_{1p}\,{}_+f_p-c'_{2p}\,{}_-f_p,
$$
$$
{}^-f_p=-c'{}^*_{2p}\,{}_+f_p+c'_{1p}\,{}_-f_p. \eqno(8)
$$

Now we have to classify solutions as in- and out-states. Our choice is [1]
$$
{}^-\psi_p\equiv{}^-\psi_{p\:out}={}_+f_p,
\qquad{}^+\psi_p\equiv{}^+\psi_{p\:out}={}^+f_p,
$$
$$
{}_-\psi_p\equiv{}_-\psi_{p\:in}=-\frac{c'_{2p}}{c'{}^*_{2p}}{}_-\tilde\psi_p,
\quad{}_-\tilde\psi_p={}_-f_p,\quad{}_+\psi_p\equiv{}_+\psi_{p\:in}={}^-f_p.
\eqno(9)
$$
Here the $\pm$ indexes in front of $\psi$-functions indicate the sign of
frequencies.

The heuristic argument in favor of this choice was based on the fact
that the description of a scattering process in terms of pure states
(unlimited vectors) is only a way to a more realistic description by
 means of wave packets. For the wave packets the field is effectively
switched off, when they leave the field region. Consider, for example, the
process described by ${}_+f_p$, see the first relation in (6). Initially
we have the antiparticle current with amplitude $c'_{1p}$ moving from the
region  of large positive $x_3$ towards the field region near $x_3=0$.
In the opposite direction from the region of large negative $x_3$ moves
the particle current with unity amplitude. It annihilates completely
in the field region. As a result we have the diminished antiparticle
current reflected from the barrier. So for $t\to\infty$ there is only
antiparticle packet, i.e. ${}_+f_p={}^-\psi_p$. We would like to
remind you here that the
momentum of the negative-frequency wave function is opposite to the
antiparticle  velocity.

In terms of in- and out-states the relations (6) and (8) take the form
$$
{}_+\psi_p=c_{1p}\,{}^+\psi_p+c_{2p}\,{}^-\psi_p,
$$
$$
{}_-\psi_p=c^*_{2p}\,{}^+\psi_p+c^*_{1p}\,{}^-\psi_p; \eqno(10)
$$
$$
{}^+\psi_p=c^*_{1p}\,{}_+\psi_p-c_{2p}\,{}_-\psi_p,
$$
$$
{}^-\psi_p=-c^*_{2p}\,{}_+\psi_p+c_{1p}\,{}_-\psi_p;  \eqno(11)
$$
$$
|c_{1p}|^2-|c_{2p}|^2=1,\qquad c_{1p}=-\frac{c'_{1p}}{c'_{2p}},\qquad
c_{2p}=\frac1{c'_{2p}}.\eqno(12)
$$
We note that in [1] the present $f-$functions were denoted as $\psi$
and it was explained that after using the transformation indicated in (9)
and (12) we get relations (10-11) with (new) $\psi$-functions. The latter
relations coincide with those for the non-stationary solutions.
 Just on the bases
of these relations the $S-$matrix formalism is build [1,5].

In terms of $\psi-$functions the field operator $\Psi$ has the usual form
$$
\Psi=\sum_p(a_{p\:in}\,{}_+\psi_p+b^{\dagger}_{p\:in}\,{}_-\psi_p)=
\sum_p(a_{p\:out}\,{}^+\psi_p+b^{\dagger}_{p\:out}\,{}^-\psi_p).\eqno(13)
$$
Here $a_{p\:in}$ is the destruction operator of a particle in the state
${}_+\psi_p$ and $b^{\dagger}_{p\:in}$ is the creation operator of an
antiparticle in the state ${}_-\psi_p$. The sum is over all $p$ including $p$
in the Klein region.

In paper [9] the field operator is written as
$$
\Psi=\sum_k[a_{k\:in}p_{k\:in}(x)+b^{\dagger}_{k\:in}n_{k\:in}(x)]=
\sum_k[a_{k\:out}p_{k\:out}(x)+b^{\dagger}_{k\:out}n_{k\:out}(x)],\eqno(14)
$$
see (31a), (31b) in [9]. Disregarding here the normalization factors, the
connections to our $\psi-$functions are
$$
p_{k\:in}={}^-\psi_k,\; n_{k\:in}={}^+\psi_k;\; p_{k\:out}={}_-\psi_k,\;
n_{k\:out}={}_+\psi_k.
               \eqno(15)
$$
The authors of [9] name their $p_{k\:in}$-function the state of incoming
particle on the grounds that there are no other waves for $x_3\to-\infty$.
In our nomenclature  this is the outgoing negative-frequency state as
explained above. So the disagreement is in the sign of frequencies and in the
in- and out- labeling. Authors of subsequent papers [10,11] accepted the
classification of [9].

Another argument in favor of our choice (9) is as follows [1]. In the case
of a constant electric field we can work either with stationary or
 non-stationary solutions. For the latter the classification of states
is obvious. The experiment should be described in terms of wave packets and
 using
stationary or non-stationary solutions should produce the same result.
This agrees with the fact that in both cases the relations (10) and (11)
have the same form with the same coefficients $c_{1p}$, $c_{2p}$. By the
way it is shown in [1] why the strict $S-$matrix formulation is possible
despite the fact that the constant field is not turned off for $t\to\pm\infty$.
The reason is that for the given set of
quantum numbers $p$ the formation length for
pair production is finite. Outside this length the field does not create
pairs and does not prevent the $S-$matrix formulation as does any field that
 does not create pairs.

 Now we mention briefly that in the scattering region $\pi^0(\pm)>m_{\perp}$, 
 we may write 
 $$
 {}^+\psi(x_3|+)\propto{}_+f_p,\quad {}_+\psi(x_3|-)\propto{}_-f_p,\quad
 {}_+\psi(x_3|+)\propto{}^+f_p,\quad {}^+\psi(x_3|-)\propto{}^-f_p \eqno(9')
 $$
 instead of (9).

 \section{Orthonormalization of wave functions}

The choice (9) assumes that ${}_+\psi$ and ${}_-\psi$ and also ${}^+\psi$ and
${}^-\psi$ are orthogonal. We shall show in this section that this is so.
 The Klein- Gordon equation
has the form
$$
[\frac{d^2}{dx_3^2}-2eA^0p^0+e^2A_0^2+p_0^2-m_{\perp}^2]f_p(x_3)
e^{i[p_1x_1+p_2x_2-p^0t]}=0. \eqno(16)
$$
 The $J^0$ component of a transition current is given by the expression
$$
J^0(\psi',\psi)=i[\psi'{}^*D_0\psi-(D_0\psi')^* \psi],
\quad D_0=\frac{\partial}{\partial t}-ieA_0. \eqno(17)
$$
For our potential $A^0$ and functions $f_{p'}$, $f_p$ we have
$$
J^0(f',f)=[p^0+p'{}^0-2eA^0(x_3)]f^*_{p'}f_p=[\pi^0(x_3)+\pi'{}^0(x_3)]
                                 f^*_{p'}f_p.                       \eqno(18)
$$
We consider also the $J_3$ component of the transition current
$$
J_3(f',f)= -if^*_{p'}\buildrel\leftrightarrow\over{\frac{\partial}
{\partial x_3}}f_p\equiv-i[f^*_{p'}\frac{\partial}{\partial x_3}f_p-
f_p\frac{\partial}{\partial x_3}f^*_{p'}]    \eqno(19)
$$
and calculate its derivatives over $x_3$ using (16)
$$
\frac{d}{dx_3}J_3=i(p^0-p'{}^0)J^0.  \eqno(20)
$$
From here we have
$$
\int\limits_{-L_d}^{L_u}dx_3J^0(f',f)=\frac{i}{p'{}^0-p^0}
[\left.J_3\right|_{x_3=L_u}-\left.J_3\right|_{x_3=-L_d}].    \eqno(21)
$$
 For $f'={}_+f_{p'}$, $f={}_+f_p$ from (5) and (19) we obtain
$$
\left.J_3({}_+f_{p'},{}_+f_p)\right|_{x_3\to-\infty}=
\frac{\pi_3(-)+\pi'_3(-)}{\sqrt{4\pi'_3(-)\pi_3(-)}}
\exp[i(\pi_3(-)-\pi'_3(-))x_3].     \eqno(22)
$$
Similarly, using in addition the first relation in (6), we find
$$
\left.J_3({}_+f_{p'},{}_+f_p)\right|_{x_3\to\infty}=
\frac{1}{\sqrt{4\pi_3(+)\pi'_3(+)}}
\{c'{}^*_{1p'}c'_{1p}[\pi_3(+)+\pi'_3(+)]\exp[i(\pi_3(+)-\pi'_3(+))x_3]
$$
$$
-c'{}^*_{2p'}c'_{2p}[\pi'_3(+)+\pi_3(+)]\exp[-i(\pi_3(+)-\pi'_3(+))x_3]+
$$
$$
 c'{}^*_{1p'}c'_{2p}[\pi'_3(+)-\pi_3(+)]\exp[-i(\pi_3(+)+\pi'_3(+))x_3]
$$
$$
+c'{}^*_{2p'}c'_{1p}[\pi_3(+)-\pi'_3(+)]\exp[i(\pi_3(+)+\pi'_3(+))x_3]\}.
                                                                 \eqno (23)
$$
For $x_3\to\infty$ the last two terms with factors
$\exp[\pm i(\pi_3(+)+\pi'_3(+))x_3]$  can be neglected; in the first two
terms we can put $p^0=p'{}^0$  everywhere except  in
$\exp[\pm i(\pi_3(+)-\pi'_3(+))x_3]$. Treating the right hand side of (22)
in a similar manner we get for (21)
$$
\int\limits_{-L_d}^{L_u}dx_3J^0({}_+f_{p'},{}_+f_p)=\frac{i}{p'{}^0-p^0}
\{|c'_{1p}|^2\exp[i(\pi_3(+)-\pi'_3(+))L_u]-
$$
$$
  |c'_{2p}|^2\exp[-i(\pi_3(+)-\pi'_3(+))L_u]-\exp[-i(\pi_3(-)-\pi'_2(-))L_d]\}.
                                                                   \eqno(24)
$$
Now according to (2)
$$
\pi'_3{}^2-\pi_3^2=\pi'_0{}^2-\pi_0^2=(\pi'{}^0+\pi^0)(p'{}^0-p^0).  \eqno(25)
$$
From here we have
$$
(\pi_3(+)-\pi'_3(+))L_u=-\frac{\pi'{}^0(+)+\pi^0(+)}
{\pi'_3(+)+\pi_3(+)}(p'{}^0-p^0)L_u.                              \eqno(26)
$$
Similarly for the phase of the expression on the right hand side of (22)
we get for $x_3=-L_d$
$$
-[\pi_3(-)-\pi'_3(-)]L_d=\frac{\pi'{}^0(-)+\pi^0(-)}{\pi'_3(-)+\pi_3(-)}
(p'{}^0-p^0)L_d.                                                  \eqno(27)
$$
We see that the expressions in (26) and (27) are of the same sign due to (4).
Using the freedom in choosing $L_d$ and $L_u$ we can make these phases equal.
Then taking into account also  relation (7), we get from (24)
$$
\int dx_3J^0({}_+f_{p'},{}_+f_p)= -|c'_{2p}|^22\pi\delta(p'{}^0-p^0).\eqno(28)
$$
So for ${}_+f_p$ normalized as in (5) we have (28). The minus sign on the right
hand side of (28) should be expected because ${}_+f={}^-\psi$ is the
negative-frequency solution of the Klein-Gordon equation. The appearance of
$|c'_{2p}|^2$ and not $|c'_{1p}|^2$ should also be expected: the density of
 particle current is subtracted from the density of antiparticle current,
$|c'_{1p}|^2-1=|c'_{2p}|$. (In the scattering region $\pi^0(\pm)>m_{\perp}$
we have to replace $-|c'_{2p}|^2$  by $|c'_{1p}|^2$ in the r.h.s. of (28).)

Now we are in a position to show that ${}^+\psi$ and ${}^-\psi$ are orthogonal.
From (19), (8) and (5) we have
\setcounter{equation}{28}
\begin{gather}
J_3({}^+f_{p'},{}_+f_p)=[c'_{1p'}\:{}_+f^*_{p'}-c'{}^*_{2p'}\:
{}_-f^*_{p'}](-i\buildrel\leftrightarrow\over{\frac{\partial}{\partial x_3}}
)\left.{}_+f_p\right|_{x_3\to-\infty}= \\  \notag
\frac{1}{\sqrt{4\pi'_3(-)\pi_3(-)}}\{ c'_{1p}[\pi'_3(-)+\pi_3(-)]
\exp[i(\pi_3(-)-\pi'_3(-))x_3] \\               \notag
-c'{}^*_{2p'}[\pi_3(-)-\pi'_3(-)]
\exp[i(\pi_3(-)+\pi'_3(-))x_3]\}.
\end{gather}
Taking into account the remarks after eq.(23) we can write
$$
\left.J_3({}^+f_{p'},{}_+f_p)\right|_{x_3=-L_d}=c'_{1p}e^
{-i[\pi_3(-)-\pi'_3(-)]L_d}=
$$
$$
c'_{1p}\exp\{i\frac{\pi'{}^0(-)+\pi^0(-)}{\pi'_3(-)+
\pi_3(-)}(p'{}^0-p^0)L_d\},  \eqno(30)
$$
where eq.(27) was used. Similarly we find
$$
J_3({}^+f_{p'},{}_+f_p)=\left.{}^+f^*_{p'}(-i\buildrel\leftrightarrow\over{
\frac{\partial}{\partial x_3})}[c'_{1p}{}^+f_p+c'_{2p}{}^-f_p)\right
|_{x_3=L_u}=
$$
$$
c'_{1p}\exp\{-i\frac{\pi'{}^0(+)+\pi^0(+)}{\pi'_3(+)+\pi_3(+)}
(p'{}^0-p^0)L_u\}.    \eqno(31)
$$
In the considered (Klein) region the exponents in (30) and (31) are of the
same sign, see (4).
Now we see that with the same adjustment of $L_d$ or $L_u$ as in deriving (28),
we have the cancellation of terms on the right hand side of (21),
i.e. ${}^+f={}^+\psi$ and ${_+f={}^-\psi}$ are orthogonal.

In the region $\pi^0(\pm)>m_{\perp}$ the arguments in favor of orthogonality 
of ${}_+\psi(x_3|+)$ and ${}_+\psi(x_3|-)$ are as follows. Having in view
eq. (21), we consider 
$$
J_3({}_+\psi_{p'}(x_3|+),{}_+\psi_{p'}(x_3|-))\propto
J_3({}^+f_{p'},{}_-f_{p})
$$
and evaluate it for $x_3\to- L_d$ with the help of the first eq. in (8),
and for $x_3\to L_u$ with the help of the second eq. in (6). Using the same
reasoning as in Klein region, we end up with the expression
$$
\int\limits_{-L_d}^{L_u}dx_3J^0({}_+\psi_{p'}(x_3|+),{}_+\psi_p(x_3|-))\propto
 c'{}^*_{2p}J_3({}^+f_{p'}(L_u),{}^+f_{p}(L_u))+
c'{}^*_{2p'}J_3({}_-f_{p'}(-L_d),{}_-f_{p}(-L_d)).    \eqno(24')
$$
By the arguments, given in connection with eqs. (30) and (31), this expression
is equal to zero (note also that $J_3({}^+f_{p},{}^+f_{p})=-
J_3({}_-f_{p},{}_-f_{p})$). For the step potential this result can be proved
by a straightforward calculation of the l.h.s. of (24') for $L_u, L_d=\infty$.
\section{Solvable potential}

 For the potential
$$
A^0(x_3)=-\tanh kx_3 \eqno(32)
$$
the solutions of the Klein-Gordon equation are known [1]
$$
{}_-f_p(x_3)=\frac{1}{\sqrt{2|\pi_3(-)|}}(-z)^{-i\mu}(1-z)^{\lambda}
F(-i\mu-i\nu+\lambda,-i\mu+i\nu+\lambda;-2i\mu+1;z),
$$
$$
-z=e^{2kx_3},\: \pi_3(-)=2k\mu,\: \pi_3(+)=2k\nu,\:\pi^0(\pm)=p^0\pm ea,\:
\lambda=\frac12+\tilde\lambda,\:\tilde\lambda=\sqrt{\frac14
-\left(\frac{ea}{k}\right)^2}.                                         \eqno(33)
$$
Here $F(\alpha,\beta;\gamma;z)$ is the hypergeometric function. $\pi_3(\pm)$
are real in the Klein region. Three other solutions with quantum numbers
$p$ can be obtained from (33) by employing the discrete symmetry of the
Klein-Gordon equation [12]. Thus
 ${}_+f_p(x_3)$ can be obtained from
 ${}_-f_p(x_3)$ by substitution $\mu\to-\mu$,
$$
{}_+f_p(x_3)=\frac{1}{\sqrt{2|\pi_3(-)|}}(-z)^{i\mu}(1-z)^{\lambda}
F(i\mu-i\nu+\lambda,i\mu+i\nu+\lambda;2i\mu+1;z), \eqno(34)
$$
${}^+f_p(x_3)$ (${}^-f_p(x_3)$) can be obtained from ${}_-f_p(x_3)$
(${}_+f_p(x_3)$) by substitutions (not changing the Klein-Gordon equation (16))
$x_3\to-x_3, a\to-a, \mu\leftrightarrow\nu, \pi^0(-)\leftrightarrow\pi^0(+)$:
$$
{}^+f_p(x_3)=\frac{1}{\sqrt{2|\pi_3(+)|}}(-z)^{i\nu}(1-z^{-1})^{\lambda}
F(-i\mu-i\nu+\lambda,i\mu-i\nu+\lambda;-2i\nu+1;z^{-1}), \eqno(35)
$$
$$
{}^-f_p(x_3)=\frac{1}{\sqrt{2|\pi_3(+)|}}(-z)^{-i\nu}(1-z^{-1})^{\lambda}
F(-i\mu+i\nu+\lambda,i\mu+i\nu+\lambda;2i\nu+1;z^{-1}), \eqno(36)
$$
The coefficients $c'_{1p}, c'_{2p}$ in (6) and (8) have the form
$$
c'_{1p}=\sqrt{\frac{\pi_3(+)}{\pi_3(-)}}\frac{\Gamma(2i\mu+1)\Gamma(2i\nu)}
{\Gamma(i\mu+i\nu+\lambda)\Gamma(i\mu+i\nu+1-\lambda)},
$$                      
$$
c'_{2p}=\sqrt{\frac{\pi_3(+)}{\pi_3(-)}}\frac{\Gamma(2i\mu+1)\Gamma(-2i\nu)}
{\Gamma(i\mu-i\nu+\lambda)\Gamma(i\mu-i\nu+1-\lambda)}. \eqno(37)
$$
Two special cases,-- the step potential ($k\to\infty$ in (32)) and constant
electric field ($k\to 0,\quad a\to\infty, \quad ak=E={\rm Const}$) are of
 particular interest. We consider them separately.
\section{Step potential}

The potential (32) takes the form
$$
A^0(x_3)= a[\theta(-x_3)-\theta(x_3)].\eqno(38)
$$
From (37) we get $(\lambda\to1-(\frac{ea}{k})^2)$
$$
c'_{1p}=\frac{\pi_3(+)+\pi_3(-)}{2\sqrt{\pi_3(+)\pi_3(-)}},\quad
c'_{2p}=\frac{\pi_3(+)-\pi_3(-)}{2\sqrt{\pi_3(+)\pi_3(-)}}. \eqno(39)
$$
The asymptotic forms (5) become exact solutions up to $x_3=0$. From relations
 (6) it follows                                                  x
$$
{}_+f_p=\frac{1}{\sqrt{2\pi_3(-)}}e^{i\pi_3(-)x_3}\theta (-x_3)+
\frac{1}{\sqrt{2\pi_3(+)}}[c'_{1p}
e^{i\pi_3(+)x_3}+c'_{2p}e^{-i\pi_3(+)x_3}]\theta(x_3),
$$
$$
{}_-f_p=\frac{1}{\sqrt{2\pi_3(-)}}e^{-i\pi_3(-)x_3}\theta (-x_3)+
\frac{1}{\sqrt{2\pi_3(+)}}[c'_{2p}
e^{i\pi_3(+)x_3}+c'_{1p}e^{-i\pi_3(+)x_3}]\theta(x_3),
 \eqno(40)
$$
Similarly, from (8) we have
$$
{}^+f_p=\frac{1}{\sqrt{2\pi_3(-)}}[c'_{1p}e^{i\pi_3(-)x_3}
-c'_{2p}e^{-i\pi_3(-)x_3}]\theta(-x_3)+\frac{1}{\sqrt {2\pi_3(+)}}
e^{i\pi_3(+)x_3}\theta(x_3),
$$
$$
{}^-f_p=\frac{1}{\sqrt{2\pi_3(-)}}[-c'_{2p}e^{i\pi_3(-)x_3}
+c'_{1p}e^{-i\pi_3(-)x_3}]\theta(-x_3)+\frac{1}{\sqrt {2\pi_3(+)}}
e^{-i\pi_3(+)x_3}\theta(x_3),
  \eqno(41)
$$
With the help of relations
$$
\int\limits_{-\infty}^0dze^{i(k-i\epsilon)z}=
\frac{1}{i(k-i\epsilon)}=\pi\delta(k)-iP\frac1k,\quad \epsilon\to+0,  \eqno(42)
$$
$$
\int\limits_0^{\infty}dze^{i(k+i\epsilon)z}=
\frac{i}{(k+i\epsilon)}=\pi\delta(k)+iP\frac1k,\quad \epsilon\to+0,  \eqno(43)
$$
where $P$ means the principal value, the integral of $J^0$ on the left hand
side of (21) can be evaluated directly. So with the help of (18) and the
first relation in (40) we find
$$
\int\limits_{-\infty}^0dx_3J^0({}_+f_{p'},{}_+f_p)=
\frac{\pi^0(-)+\pi'{}^0(-)}{\sqrt{4\pi_3(-)\pi'_3(-)}}\{\pi\delta[\pi_3(-)
-\pi'_3(-)]-iP\frac{1}{\pi_3(-)-\pi'_3(-)}\}.  \eqno(44)
$$
Similarly, we get
$$
\int\limits_0^{\infty}dx_3J^0({}_+f_{p'},{}_+f_p)=
$$
$$
\frac{\pi^0(+)+\pi'{}^0(+)}{\sqrt{4\pi_3(+)\pi'_3(+)}}\int\limits_0^{\infty}
dx_3[c'_{1p'}e^{-i\pi'_3(+)x_3}+c'_{2p'}e^{i\pi'_3(+)x_3}]
[c'_{1p}e^{i\pi_3(+)x_3}+c'_{2p}e^{-i\pi_3(+)x_3}]=
$$
$$
\frac{\pi^0(+)+\pi'{}^0(+)}
{\sqrt{4\pi_3(+)\pi'_3(+)}}\{c'_{1p'}c'_{1p}[\pi\delta(\pi_3(+)-\pi'_3(+))+
iP\frac{1}{\pi_3(+)-\pi'_3(+)}]-
ic'_{1p'}c'_{2p}P\frac{1}{\pi'_3(+)+\pi_3(+)}+
$$
$$
ic'_{2p'}c'_{1p}P\frac{1}{\pi'_3(+)+\pi_3(+)} +c'_{2p'}c'_{2p}[\pi\delta
(\pi'_3(+)-\pi_3(+))+iP\frac{1}{\pi'_3(+)-\pi_3(+)}]\}.
 \eqno(45)
$$
Here $\delta-$functions with nonzero argument are dropped. Using (25)
the term with $\delta-$functions in (44) can be written as
$$
\pi\delta(p'{}^0-p^0). \eqno(46)
$$
 Similarly, terms with $\delta-$function in (45)   can be simplified
$$\frac{\pi^0(+)}{\pi_3(+)}(c'{}^2_{1p}+c'{}^2_{2p})\frac{\pi_3(+)}
{|\pi^0(+)|}\pi\delta(p'{}^0-p^0)=\frac{\pi^0(+)}{|\pi^0(+)|}
(c'{}^2_{1p}+c'{}^2_{2p})\pi\delta(p'{}^0-p^0).    \eqno(47)
$$
On account of (7) the sum of (46) and (47) gives the right hand side of (28)
for $\pi^0(+)=-|\pi^0(+)|$.

Now we verify that in the sum of (44) and (45) the terms with $iP$ are
cancelled out. Using (25) the term with $iP$ in (44) can be written in the form
$$
\frac{\pi_3(-)+\pi'_3(-)}{\sqrt{4\pi_3(-)\pi'_3(-)}}\frac1{p'{}^0-p^0}. \eqno(48)
$$
Similarly, the first term with $iP$ in (45) acquires the form
$$
\frac{c'_{1p'}c'_{1p}}{\sqrt{4\pi_3(+)\pi'_3(+)}}\frac{\pi_3(+)+\pi'_3(+)}
{p'{}^0-p^0}.  \eqno(49)
$$
In a similar manner the fourth term with $iP$ in (45) becomes
$$
-\frac{c'_{2p'}c'_{2p}}{\sqrt{4\pi_3(+)\pi'_3(+)}}\frac{\pi_3(+)+\pi'_3(+)}
{p'{}^0-p^0}.   \eqno(50)
$$
From definitions of $c'_{1p}$ and $c'_{2p}$ in (39) we have
$$
c'_{1p'}c'_{1p}-c'_{2p'}c'_{2p}=\frac{\pi'_3(+)\pi_3(-)+\pi'_3(-)\pi_3(+)}
{\sqrt{4\pi'_3(+)\pi'_3(-)\pi_3(+)\pi_3(-)}}  , \eqno(51)
$$
$$
c'_{2p'}c'_{1p}-c'_{2p}c'_{1p'}=\frac{\pi'_3(+)\pi_3(-)-\pi'_3(-)\pi_3(+)}
{\sqrt{4\pi'_3(+)\pi'_3(-)\pi_3(+)\pi_3(-)}}  .\eqno(51a)
$$
Taking this into account, the sum of (49) and (50) acquires the form
$$
\frac{1}{4\sqrt{\pi_3(-)\pi'_3(-)}}\frac{1}{p^0-p'{}^0}\{\pi_3(-)+
\frac{\pi'_3(-)\pi_3(+)}{\pi'_3(+)}+\frac{\pi'_3(+)\pi_3(-)}{\pi_3(+)}+
\pi'_3(-)\}.    \eqno(52)
$$
In the same way we find the sum of the second and the third terms in (45)
$$
\frac{1}{4\sqrt{\pi_3(-)\pi'_3(-)}}\frac{1}{p^0-p'{}^0}\{\pi_3(-)-
\frac{\pi'_3(-)\pi_3(+)}{\pi'_3(+)}-\frac{\pi'_3(+)\pi_3(-)}{\pi_3(+)}+
\pi'_3(-)\}. \eqno(53)
$$
Now it is seen that the sum of (52) and (53) is equal to (48) with the
 opposite sign, q.e.d.

Now  we show that ${}^-\psi$ and ${}^+\psi$ are orthogonal. The contribution
from the negative $x_3$ to the integral on the left hand side of (21) is,
see (18),
$$
\int\limits_{-\infty}^0dx_3J^0({}^+f_{p'},{}_+f_p)=[\pi'{}^0(-)+\pi^0(-)]
\int\limits_{-\infty}^0dx_3{}^+f^*_{p'}{}_+f_p= \eqno(54)
$$
$$
\frac{\pi'{}^0(-)+\pi^0(-)}
{\sqrt{4\pi_3(-)\pi'_3(-)}}
\{ c'_{1p'}[\pi\delta(\pi_3(-)-\pi'_3(-))-iP\frac{1}{\pi_3(-)-\pi'_3(-)}]+
c'_{2p'}iP\frac{1}{\pi_3(-)+\pi'_3(-)}\}.   \eqno(55)
$$
In the last equation the first relation (8) and eq.(42) were used. Similarly
we obtain
$$
\int\limits_0^{\infty}dx_3J^0({}^+f_{p'},{}_+f_)=
$$
$$
\frac{\pi'{}^0(+)+\pi^0(+)}
{\sqrt{4\pi_3(+)\pi'_3(+)}}
\{ c'_{1p}[\pi\delta(\pi_3(+)-\pi'_3(+))+iP\frac{1}{\pi_3(+)-\pi'_3(+)}]-
c'_{2p}iP\frac{1}{\pi_3(+)+\pi'_3(+)}\}.  \eqno(56)
$$
Using (25) we write the term with $\delta-$function in (55) as
$$
c'_{1p}\pi\delta(p^0-p'{}^0). \eqno(57)
$$
 Similar term in (56) gives (57) with the opposite sign.

Now we check that the sum of terms with $iP$ in (55) and (56) is also zero.
The first term with $iP$ in (55) on account of (25) and (39) is
$$
c'_{1p'}\frac{\pi'{}^0(-)+\pi^0(-)}{\sqrt{4\pi_3(-)\pi'_3(-)}}\frac
{1}{\pi'_3(-)-\pi_3(-)}=\frac{[\pi'_3(+)+\pi'_3(-)][\pi_3(-)+\pi'_3(-])}
{4\pi'_3(-)\sqrt{\pi_3(-)\pi'_3(+)}}\frac{1}{p'{}^0-p^0}. \eqno(58)
$$
In the same manner we get the second term
$$
c'_{2p'}\frac{\pi'{}^0(-)+\pi^0(-)}{\sqrt{4\pi_3(-)\pi'_3(-)}}\frac
{1}{\pi'_3(-)+\pi_3(-)}=\frac{[\pi'_3(+)-\pi'_3(-)][\pi'_3(-)-\pi_3(-])}
{4\pi'_3(-)\sqrt{\pi_3(-)\pi'_3(+)}}\frac{1}{p'{}^0-p^0}.
 \eqno(59)
$$
The sum of these two terms is
$$
\frac{\pi'_3(+)+\pi_3(-)}{2\sqrt{\pi_3(-)\pi'_3(+)}}\frac{1}{p'{}^0)-p^0}.
  \eqno(60)
$$
Similar calculation of terms with $iP$ in (56) produces (60) with the
 opposite sign. So the sum of all terms is zero.

\section{Constant electric field}

In this Section we obtain the propagator for the scalar particle in a
 constant electric field and show that in terms of our in- and out-states
it has the form dictated by the general theory. The vector-potential (32)
reduces to
 $$
A_{\mu}=\delta_{\mu0}Ex_3,\quad A^0=-A_0.    \eqno(61)
$$
With this potential the Klein-Gordon equation (16) takes on the form
$$
[\frac{d^2}{dZ^2}+\frac{Z^2}4-\frac{\lambda}2]f(x_3)=0, \eqno(62)
$$
$$
Z=\sqrt{2|eE|}(x_3+\frac{p^0}{eE})=-\sqrt{\frac{2}{|eE|}}\pi^0(x_3),
\quad \lambda=\frac{m_{\perp}^2}{|eE|}.
 \eqno(63)
$$
The solutions, normalized on unity current,  are, see eq.(8.2.5) in [13],
$$
{}^-\psi_p={}_+f_p=c_pD_{\nu}(-e^{\frac{i\pi}{4}}Z),
{}_-\psi_p={}_-f_p=c_pD_{\nu^*}(-e^{-\frac{i\pi}{4}}Z),
$$
$$
{}_+\psi_p={}^-f_p=c_pD_{\nu}(e^{\frac{i\pi}{4}}Z),
{}^+\psi_p={}^+f_p=c_pD_{\nu^*}(e^{-\frac{i\pi}{4}}Z),\eqno(64)
$$
$$
\nu=\frac{i\lambda}{2}-\frac12, \quad c_p=(2|eE|)^{-\frac14}
\exp[\frac{\pi\lambda}{8}],\quad p=p_1,p_2,p^0.    \eqno(65)
$$
The normalization factor $c_p$ is found with the help of Wronskian
$$
D_{-1-\nu}(-ix)\buildrel\leftrightarrow\over{\frac{d}{dx}}D_{\nu}(x)=
-i\exp[\frac{i\pi}{2}\nu],  \eqno(66)
$$
    which follows from eqs.(8.2.10) and (8.2.8) in [13]. Now in the relations
(6) and (8)
$$
c'_{1p}=\frac{\sqrt{2\pi}}{\Gamma(\frac{1-i\lambda}{2})}\exp[\frac{\pi}{4}
(\lambda-i)],\quad   c'_{2p}=i\exp[\frac{\pi}{2}\lambda],   \eqno(67)
$$
and in relations (10) and (11)
$$
c_{1p}=-\frac{c'_{1p}}{c'_{2p}}=\frac{\sqrt{2\pi}}{\Gamma(\frac{1-i\lambda}{2})}
\exp[-\frac{\pi}{4}
(\lambda-i)],\quad   c_{2p}=\frac{1}{c'_{2p}}=-i\exp[-\frac{\pi}{2}\lambda],     \eqno(68)
$$

We note that ${}_-\psi$ and ${}_-\tilde\psi$ in (9) coincide in this case.
According to (28) to normalize $\psi_p$ on $\pm2\pi\delta(p^0-p'{}^0)$
we have to replace $c_p$ in (64) and (65) by
$$
\frac{c_p}{|c'_{2p}|}=(2|eE|)^{-\frac14}\exp[-\frac{3\pi\lambda}{8}]. \eqno(69)
$$
 Thus we may assume that $\psi$-functions in (10) and (11) are normalized
in this way. The same relations hold for non-stationary states. The important
thing is that the relations (10) and (11) constitute all the necessary
ingredients for $S-$matrix theory [1,5].

We note now that the relation (18) takes on the form
$$
J^0(f_{p'},f_p)=-\sqrt{\frac{|eE|}{2}}(Z+Z')f^*_{p'}f_p,
\quad Z'=\sqrt{2|eE|}(x_3+\frac{p'{}^0}{eE}),\quad e=-|e|.   \eqno(70)
$$
On the other hand, using (62) we find
$$
\frac{d}{dZ}[f^*_{p'}\buildrel\leftrightarrow\over{\frac{d}{dZ}}f_p]=
-\frac{1}{\sqrt{8|eE|}}(p'{}^0-p^0)(Z+Z')f^*_{p'}f_p.     \eqno(71)
$$
Now it easy to verify that relations (19), (21) and (28) remain valid
as well as the orthogonality of ${}_+\psi$ and ${}_-\psi$ and also of
${}^+\psi$ and ${}^-\psi$.

 Solutions (64), in which the factor depending on $x_1, x_2, t$ is
dropped for brevity, are characterized by the quantum number $p^0$
(and also $p_1$ and $p_2$). If instead of vector-potential (61) we use
in the Klein- Gordon equation the vector-potential
$$
A'_{\mu}=-\delta_{\mu3}Ex^-=A_{\mu}-\frac{\partial}{\partial x^{\mu}}\eta,
\:x^{\pm}=t\pm x_3,\:\eta=E(tx_3-\frac{x_3^2}{2}),    \eqno(72)
$$
we obtain the solutions $\psi'_{p^-}$ characterized by the quantum number
$p^-$, see [14,15]. Of course, we can go back to the potential (61) and
obtain
$$
\psi_{p^-}(x|A)\equiv\psi_{p^-}=e^{ie\eta}\psi'_{p^-},\quad
\psi'_{p^-}\equiv\psi'_{p^-}(x|A').    \eqno(73)
$$
Making modifications due to the present assumption $e=-|e|$, we have from
the results in [14, 15]
$$
{}^+\psi'_{p^-}(x)=\frac{1}{(4|eE|)^{\frac14}}\exp[-\frac{i}{2}p^-x^++
\frac{i}{4}eE(x^-)^2+\nu^*\ln z],
\:z=\frac{\pi^-}{\sqrt{|eE|}},\:\pi^-=p^--eEx^-.               \eqno(74)
$$
The factor depending on $x_1, x_2$ is dropped for brevity.
So this is the positive-frequency out-state. To corroborate this we may
add to the arguments in [14, 15] the following physical justification. The
classical particle with the negative charge starts from the region with
large negative $x_3$, is slowed down and is reflected back to where it comes.
Its kinetic momentum $\pi_3$ is negative and grows in magnitude for
$t\to\infty$. Hence
$$
\pi^-=\left.(\pi^0-\pi_3)\right|_{t\to\infty}\to\infty.  \eqno(75)
$$
Going back to the quantum state (74) we note that
 the particles (antiparticles) are in the region where $\pi^->0$
($\pi^-<0$). Large $\pi^-$ indicates that we are far from the region where
pairs are created ($\pi^-\approx 0$) .
In the region where $\pi^-<0$ the wave function (74) must be small for
small probability of pair production. Thus in this region
$$
z=e^{-i\pi}(-z),\quad z<0.                                                 \eqno(76)
$$
Similarly,
$$
{}_-\psi'_{p^-}(x)=\frac{1}{(4|eE|)^{\frac14}}\exp[-\frac{i}{2}p^-x^++
\frac i4eE(x^-)^2+\nu^*\ln(-z)].                                     \eqno(77)
$$
For positive $z$ in (77) we have
$$
-z=e^{-i\pi}z,\quad z>0.                                                     \eqno(78)
$$
The motivation for the normalization factors in (74) is as follows.
 Introducing the operators
$$
\Pi_3=\Pi^3=-i\frac{\partial}{\partial x_3}-eA_3,\:\Pi^0=-\Pi_0=i\frac{\partial}
{\partial t}+eA_0,\: \Pi^-=\Pi^0-\Pi^3=2i\frac{\partial}{\partial x^+}+
e(A_0+A_3),                                                   \eqno(79)
$$
we find
$$
\Pi^-\psi_{p^-}=\pi^-\psi_{p^-},\quad \pi^-=p^--eEx^-.            \eqno(80)
$$
So for $\psi-$functions in (74) and (77) we obtain [14, 15]
$$
\int\limits_{-\infty}^{\infty} dx^+\,{}^+\psi^*_{p'{}^-}\,\Pi^-\,{}^+\psi_{p^-}
=2\pi\delta(p'{}^0-p^0),
\: \pi^->0,                                                          \eqno(81)
$$
$$
\int\limits_{-\infty}^{\infty} dx^+\,{}_-\psi^*_{p'{}^-}\,\Pi^-\,{}_-\psi_{p^-}
=-2\pi\delta(p'{}^0-p^0),
\: \pi^-<0.                                                          \eqno(82)
$$
These relations hold both for $\psi_{p^-}$ and for $\psi'_{p^-}$.

   The complete  set of $\psi_{p^-}$-solutions must satisfy the relations
(10) and (11). This gives
$$
{}_+\psi'_{p^-}=\theta(\pi^-)c_{1p}{}^+\psi'_{p^-},
{}^-\psi'_{p^-}=\theta(-\pi^-)c_{1p}{}_-\psi'_{p^-},
$$
$$
   \theta (x)=\left\{\begin{array}{cc}
1,\quad x>0\\
0,\quad x<0.
\end{array}\right.                                                     \eqno(83)
$$
The proper time representation of the scalar particle propagator for the
vector-potential (72) is [16,1]
$$
G(x',x|A')
=\frac{eE}{(4\pi)^2}
\exp[-i\frac{eE}{2}y_3(x'{}^-+x^-)]
\int\limits_0^{\infty}\frac{ds}{s\sinh(eEs)}\times
$$
 $$
\exp[-ism^2+i\frac{y_1^2+y_2^2}{4s}-
\frac{i}{4}(y_0^2-y_3^2)eE\coth(eEs)],\quad y=x'-x.                 \eqno(84)
 $$
Now we multiply (84) by $\exp\{-i[p_1y_1+p_2y_2-\frac12p^-y^+]\}$ and
 integrate over $y_1, y_2$ and $y^+$, see (91).
For the integral over $y^+$ we have
$$
\int\limits_{-\infty}^{\infty}dy^+\exp[\frac{i}{2}p^-y^+-\frac{i}{4}eEy^+(x^-+
x'{}^-)-\frac{i}{4}y^-y^+eE\coth(eEs)]=
$$
 $$
2\pi\delta[\frac{p^-}{2}-\frac{eE}{4}(x^-+
x'{}^-)-\frac{y^-}{4}eE\coth(eEs)].                                   \eqno(85)
 $$
In the expressions (84) and (85) the charge $e$ can have any sign. For
$e=-|e|$ we obtain for the right hand side of (85)
$$
\frac{8\pi}{|eEy^-|}\delta(\coth\tau-R)=\frac{8\pi}{|eEy^-|}
\sinh^2\tau_0\delta(\tau-
\tau_0),\quad R=\frac{\pi'{}^-+\pi^-}{\pi'{}^--\pi^-},                \eqno(86)
$$
 $$
\tau=|eE|s,\: \tau_0=\frac12\ln\frac{R+1}{R-1}=\frac12\ln\frac{\pi'{}^-}{\pi^-},
\: \pi^-=p^-+|eE|x^-,\: \pi'{}^-=p^-+|eE|x'{}^-,\: y^-=x'{}^--x^-.    \eqno(87)
$$
We note that the reversal of sign of $\pi^-$ and $\pi'{}^-$ does not change
$R$. In order to have the nonzero $\delta-$function argument, $R$ must lie in
the interval $1<R<\infty$ because $s>0$ in (84). This is possible only
in two cases
$$
0<\pi^-<\pi'{}^-,\quad{\rm i.e.} \quad x^-<x'{}^-; \eqno(88)
$$
$$
0>\pi^->\pi'{}^-,\quad{\rm i.e.} \quad  x^->x'{}^-.  \eqno(89)
$$
Taking into account that
$$
  \int\limits_{-\infty}^{\infty}dy_1\exp[\frac{i}{4s}y_1^2-ip_1y_1]=2\sqrt
{\pi s}\exp[\frac{i\pi}{4}-isp_1^2]                                     \eqno(90)
$$
and similarly for the integral over $y_2$, we get for the case (88)
$$
\int\limits_{-\infty}^{\infty}dy_1\int\limits_{-\infty}^{\infty}dy_2
\int\limits_{-\infty}^{\infty}dy^+G(x',x|A')
e^{-i[p_1y_1+p_2y_2-\frac12p^-y^+]}
$$
$$
=i(\pi^-\pi'{}^-)^{-\frac12}\exp\{-\frac{i\lambda}{2}\ln\frac{\pi'{}^-}{\pi^-}+
\frac{i}{4}eE[(x'{}^-)^2-(x^-)^2\}.                               \eqno(91)
$$
The relation $e^{\tau^0}=\sqrt{\frac{\pi'{}^-}{\pi^-}}$ is used here, see (87).
 We note also that according to (74) and (77) both
 ${}^+\psi'_{p^-}(x'){}^+\psi'{}^*_{p^-}(x)$ under
 condition
(88) and ${}_-\psi'_{p^-}(x'){}_-\psi'{^*}_{p^-}(x)$  under condition (89)
 can be written as
$$
 \frac{1}{2\sqrt{\pi^-\pi'{}^-}}\exp\{-\frac{ip^-}{2}(x'{}^+-x^+)+
\frac{i}{4}eE[(x'{}^-)^2-(x^-)^2]-\frac{i\lambda}{2}\ln\frac{\pi'{^-}}{\pi^-}\}.
                                                                \eqno(92)
$$
Making the inverse Fourier transform of (91) i.e. multiplying it by
$2^{-1}(2\pi)^{-3}\exp[i(p_1y'_1+p_2y'_2-\frac12p^-y'{}^+)]$
 and integrating over $p_1, p_2$ and $p^-$, we get
$$
G(x',x|A')=i\int\frac{dp_1dp_2dp^-}{(2\pi)^3}\left\{\begin{array}{cc}
\theta(\pi^-)\theta(\pi'{}^-){}^+\psi'_{p^-}(x')\:{}^+\psi'{}^*_{p^-}(x),
\quad  x'{}^->x^-,\\
\theta(-\pi^-)\theta(-\pi'{}^-){}_-\psi'_{p^-}(x')\:{}_-\psi'{}^*_{p^-}(x),
\quad x'{}^-<x^-.
  \end{array}\right.                                              \eqno(93)
$$
Here $\theta-$functions take care of the conditions (88) and (89).
Besides we have
$$
\theta(\pi^-)\theta(\pi'{}^-)=\theta(\pi^-), \quad x'{}^->x^-,
$$
 $$
\theta(-\pi^-)\theta(-\pi'{}^-)=\theta(-\pi^-), \quad x'{}^-<x^-.   \eqno(94)
 $$
 Taking into account (83) we obtain from (93) and (94) the sought for
representation, which for the potential (61) has the form
$$
G(x',x|A)=i\int\frac{dp_1dp_2dp^-}{(2\pi)^3c^*_{1p}}\left\{\begin{array}{cc}
{}^+\psi_{p^-}(x')\:{}_+\psi{}^*_{p^-}(x),
\quad  x'{}^->x^-,\\
{}_-\psi_{p^-}(x')\:{}^-\psi{}^*_{p^-}(x),
\quad x'{}^-<x^-.
  \end{array}\right.                                                  \eqno(95)
$$
One can verify [15] that the functions defined by the upper and lower
 lines on the right hand side of (93) (and (95)) coincide outside the light cones
$$
y^+y^-=y_0^2-y_3^2 <0.       \eqno(96)
$$
According to (73) and (74) we have
$$
{}^+\psi_{p^-}=(4|eE|)^{-\frac14}\exp[-\frac{i}{2}p^-x^++ieE(\frac{t^2}{2}-
\frac{(x^-)^2}{4})+\nu^*\ln z].         \eqno(97)
$$
${}_-\psi_{p^-}$ is obtained from (97) by substitution $z\to-z$ under the
logarithm sign.

It is instructive to go back from (93) to (84). Using the definitions of
$\pi^-$ and $\pi'{}^-$ in (87), we rewrite (92) as follows
$$
\psi'_{p^-}(x')\psi'{}^*_{p^-}(x)=
$$
$$
(4\pi'{}^-\pi^-)^{-\frac12}
\exp[-\frac{i}{2}eEy_3(x^-+x'{}^-)-\frac{i}{4}y^+(\pi'{}^-+\pi^-)
-\frac{i}{2}\lambda
\ln\frac{\pi'{}^-}{\pi^-}].                                        \eqno(98)
$$
Now we consider the first line on the right hand side of (93). The integral
over $p^-$ can be considered as an integral over $\pi^-$. If instead of
$\pi^-$ we use $\tau$
$$
\tau\equiv|eE|s=\frac12\ln\frac{\pi'{}^-}{\pi^-},\: \coth \tau=R=1+\frac{2\pi^-}
{|eE|y^-},\quad  \pi'{}^--\pi^-=|eE|y^-,                       \eqno(99)
$$
we get
$$
\int\limits_0^{\infty}\frac{d\pi^-}{\sqrt{\pi^-\pi'{}^-}}
\exp[-\frac{i}{4}y^+(\pi'{}^-+\pi^-)-
\frac{i}{2}\lambda\ln\frac{\pi'{}^-}{\pi^-}]=
$$
 $$
\int\limits_0^{\infty}\frac{d\tau}{\sinh\tau}\exp[-i\lambda\tau-
\frac{i}{4}(y_0^2-
y_3^2)|eE|\coth\tau].                                            \eqno(100)
 $$
 Finally, using
$$
\int\limits_{-\infty}^{\infty}dp_1\exp[ip_1y_1-\frac{i}{|eE|}p_1^2\tau]=
\sqrt{\frac{\pi}{is}}\exp[\frac{iy_1^2}{4s}]                     \eqno(101)
$$
and similar expression for the integral over $p_2$, we recover (84). We see that
the proper time $s$ is expressed through $\pi^-$ and $\pi'{}^-$ according
(99).

Now we note that the classification of $\psi_{p^-}$-functions was obtained in
[14,15] from the obvious classification of the non-stationary solutions
$\psi_{p_3}$ by an integral transformation
$$
\psi_{p^-}(x|A)=\int\limits_{-\infty}^{\infty}dp_3M^*(p_3,p^-)\varphi_{p_3}(x|A),
                                                                    \eqno(102)
$$
 $$
 M^*(p_3,p^-)=(2\pi|eE|)^{-\frac12}\exp\{i\frac{(p^-)^2+4p^-p_3+2p_3^2}{4eE}\}.
                                                                   \eqno(103)
 $$
Using ${}^+\varphi_{p_3}(x|A)$ in the expression on the right hand side
of (102), we get ${}^+\psi_{p^-}(x|A)$ on the left hand side and so on.
The reversed relations of (102) can be considered as the definitions of
$\varphi_{p_3}$. For the vector potential $\tilde A_{\mu}=-\delta_{\mu3}Et$,
considered in [14, 15], we have
$$
 \int\limits_{-\infty}^{\infty}dp^-M(p_3,p^-){}^+\psi_{p^-}(x|\tilde A)=
{}^+\varphi_{p_3}(x|\tilde A)=
B_pe^{ip_3x_3}D_{\nu^*}(e^{\frac{i\pi}4}T),
$$
 $$
 \int\limits_{-\infty}^{\infty}dp^-M(p_3,p^-){}_-\psi_{p^-}(x|\tilde A)=
{}_-\varphi_{p_3}(x|\tilde A)=
B_pe^{ip_3x_3}D_{\nu^*}(-e^{\frac{i\pi}4}T),
 $$
$$
B_p=(2|eE|)^{-\frac14}\exp[-\frac{\pi\lambda}{8}-\frac{i\lambda}{4}\ln2+
\frac{i3\pi}{4}],\quad T=\sqrt{2|eE|}(t-\frac{p_3}{|eE|}).           \eqno(104)
$$
By analogy we expect that
$$
\varphi_{p^0}=\int\limits_{-\infty}^{\infty}dp^-K(p^0,p^-)\psi_{p^-},
                                                                   \eqno(105)
$$
 $$
K (p^0,p^-)=(2\pi|eE|)^{-\frac12}\exp\{i\frac{(p^-)^2-4p^-p^0+2p_0^2}{4eE}\},
                                                                     \eqno(106)
 $$
where $\varphi_{p^0}$ differs from $\psi_p=\psi_{p^0}$ in (64) only by an
 inessential  phase factor, see (110). To see that this is true we use first
the relation
$$
\frac{i}{4eE}[(p^-)^2-4p^-p^0+2p_0^2]+i[-\frac{p^-x^+}{2}+\frac{eEt^2}{2}-
\frac{eE}{4}(x^-)^2]=
$$
$$
-i\frac{z^2}{4}+iz\zeta-i\frac{\zeta^2}{2}-ip^0t,
                                                                  \eqno(107)
$$
where
$$
\zeta=-\sqrt{|eE|}(x_3+\frac{p^0}{eE}),\quad z=\frac{\pi^-}{\sqrt{|eE|}},
\quad e=-|e|,                                                         \eqno(108)
$$
then formula 3.462(1) in [17], the prescriptions (76) and (78)  and the
 relations between the parabolic cylinder functions, see 8.2(6)-(8) in [13].
Then we find
$$
{}^+\varphi_{p^0}=\int\limits_{-\infty}^{\infty}dp^-K(p^0,p^-){}^+\psi_{p^-}
$$
 $$
=(2|eE|)^{-\frac14}\exp[i\frac{\pi}{8}-\frac{i\lambda}{4}\ln2
-\frac{3\pi\lambda}{8}
-ip^0t]D_{\nu^*}(e^{-i\frac{\pi}{4}}Z),                             \eqno(109)
 $$
 $$
{}_-\varphi_{p^0}=\int\limits_{-\infty}^{\infty}dp^-K(p^0,p^-){}_-\psi_{p^-}
$$
$$
=(2|eE|)^{-\frac14}\exp[i\frac{\pi}{8}-\frac{i\lambda}{4}\ln2-\frac{3\pi\lambda}{8}
-ip^0t]D_{\nu^*}(-e^{-i\frac{\pi}{4}}Z),                             \eqno(110)
$$
and similar expressions for ${}_+\varphi_{p^0}$ and
 ${}^-\varphi_{p^0}$.
Here $Z$ is the same as in (63).

Comparing these functions with the ones in (64) together with (69), we see
that $\varphi_{p^0}$ coincide with $\psi_{p^0}$ up to an inessential
phase factor. We could get rid of this factor by modifying $K(p^0,p^-)$.

We note here that
$$
\int\limits_{-\infty}^{\infty}dp^-K^*(p'{}^0,p^-)K(p^0,p^-)=\delta(p^0-p'{}^0)
                                                                  \eqno(111)
$$
and
$$
\int\limits_{-\infty}^{\infty}dp^0K^*(p^0,p'{}^-)K(p^0,p^-)=\delta(p^--p'{}^-).
                                                                   \eqno(112)
$$
So the relation (105) can be reversed
$$
\psi_{p^-}=\int\limits_{-\infty}^{\infty}dp^0K^*(p^0,p^-)\varphi_{p^0}.
                                                                   \eqno(113)
$$
Using this formula we rewrite the integrals over $p^-$ in (95) as follows
$$
\int\limits_{-\infty}^{\infty}dp^-{}^+\psi_{p^-}(x'){}_+\psi^*_{p_-}(x)=
\int\limits_{-\infty}^{\infty}dp^0{}^+\psi_{p^0}(x'){}_+\psi^*_{p_0}(x),
\:t'>t,
$$
$$
\int\limits_{-\infty}^{\infty}dp^-{}_-\psi_{p^-}(x'){}^-\psi^*_{p_-}(x)=
\int\limits_{-\infty}^{\infty}dp^0{}_-\psi_{p^0}(x'){}^-\psi^*_{p_0}(x),
\:t'<t.                                                            \eqno(114)
$$
The conditions on $t$ and $t'$ are written on account of the remark after (95).
Thus
$$
G(x',x|A)=i\int\limits_{-\infty}^{\infty}dp_1\int\limits_{-\infty}^{\infty}dp_2
\int\limits_{-\infty}^{\infty}dp^0\frac{1}{(2\pi)^3c^*_{1p}}\times
$$
 $$
\left\{\begin{array}{cc}
{}^+\psi_{p^0}(x'){}_+\psi^*_{p^0}(x),\quad t'>t,\\
{}_-\psi_{p^0}(x'){}^-\psi^*_{p^0}(x),\quad t'<t.
\end{array}\right.                                               \eqno(115)
 $$
By the way  it is clear from (115) and (73) that
$$
G(x',x|A')=e^{ie(\eta(x)-\eta(x'))}G(x',x|A).                  \eqno(116)
$$
The expression  (115) has the form dictated by the general
theory [1, 18].

Finally we note that it follows from (102) and (105) that
$$
\varphi_{p^0}=\int\limits_{-\infty}^{\infty}dp_3N(p^0,p_3)\varphi_{p_3},
                                                                   \eqno(117)
$$
where
$$
N(p^0,p_3)=\int\limits_{-\infty}^{\infty}dp^-K(p^0,p^-)M^*(p_3,p^-)=
(2\pi|eE|)^{-\frac12}\exp[-\frac{i\pi}{4}-\frac{ip^0p_3}{|eE|}].     \eqno(118)
$$
The relation (117) can be checked with the help of formula 2.11.4(7) in [19]
which can be adjusted as follows
$$
\int\limits_{-\infty}^{\infty}dx\exp[\frac12c^2xy]
D_{\varkappa}(cx)=
\frac{2\sqrt{\pi}}{c}(-1)^{\frac{\varkappa}{2}}D_{\varkappa}(\frac{cy}{\sqrt{-1}}),
\quad|{\rm phase}|\: c<\frac{3\pi}{4}.                             \eqno(119)
$$
Here $\sqrt{-1}=i$ ($\sqrt{-1}=-i$) for $\varkappa=\nu$ ($\varkappa=\nu^*$).
So we insert ${}^+\varphi_{p_3}(x|A)$ into the right hand side of (117) and take
into account that
$$
{}^+\varphi_{p_3}(x|A)=e^{-i|eE|tx_3}\:{}^+\varphi_{p_3}(x|\tilde A),\quad
-i\frac{p^0p_3}{|eE|}+p_3x_3=-\frac{i}{2}ZT-ip^0t+i|eE|tx_3,\quad
                              \eqno(120)
$$
Then we get
$$
{}^+\varphi_{p^0}(x|A)=
\int\limits_{-\infty}^{\infty}dp_3N(p^0,p_3){}^+\varphi_{p_3}(x|A)=
\frac{e^{\frac{i\pi}{8}-\frac{\pi\lambda}{8}-\frac{i\lambda}{4}\ln 2
-ip^0t}}{2\sqrt{\pi}(2|eE|)^{\frac14}}\int\limits_{-\infty}^{\infty}dT\exp[-\frac{i}{2}ZT]
 D_{\nu^*}(e^{\frac{i\pi}{4}}T).    \eqno(121)
$$
Now using (119) we obtain the right hand side of (109). For
${}_-\varphi_{p_3}(x|A)$ we proceed similarly, but use the substitution
$T=-x$ instead of $T=x$, when employing (119).

It is shown in [14, 15] that for example ${}^+\psi_{p^-}$ can be obtained
 from ${}^+\psi_{p_3}$
by changing continuously the gauge of the electric field potential. The same is true
for ${}^+\psi_{p^-}$ and ${}^+\psi_{p^0}$. For this reason these functions
 are indistinguishable and only the wave packets are observable.

 It is clear that
$N(p^0,p_3)$ and
$M(p_3,p^-)$
  have the orthogonality properties of $K(p^0,p^-)$, see (111) and (112).

\section{Acknowledgments}
 I am greatly indebted to Prof. V.L.Ginzburg, who showed me article [11],
discussed it and urged me to expound my understanding of the problem and
points of disagreement with other authors.
This work was supported in part by the Russian Foundation for Basic Research
(projects no. 00-15-96566 and 02-02-16944)

 \section*{References}
\begin{enumerate}
\item
A.I.Nikishov, Tr. Fiz. Inst. Akad. Nauk SSSR {\bf 111}, 152 (1979); J. Sov.
 Laser Res. {\bf 6},\\
 619 (1985). \\
\item  A.I.Nikishov, in {\sl Problems in Theoretical Physics},Moscow, Nauka, p.
 229, 1972 (in Russian).\\
\item   A.I.Nikishov, Teor. Mat. Fiz. {\bf 20}, 48 (1974).\\
\item A.A.Grib, S.G.Mamaev, and V.M.Mostepanenco,{\sl Vacuum Quantum Effects
  in Strong Fields} (Energoatomizdat, Moscow, 1988).\\
\item A.I.Nikishov, Tr. Fiz. Inst. Akad. Nauk SSSR  {\bf 168}, 157 (1985); \\
  in {\sl Issues in Intensive-Field Quantum Electrodynamics}, Ed. by
  V.L.Ginzburg (Nova Science, Commack, 1987).   \\
\item A.I.Nikishov, Zh.\'Eksp.Teor.Fiz. {\bf 57}, 1210 (1969) [Sov. Phys. JETP
   {\bf 30}, 660 (1970)].                          \\
\item  N.B.Narozhny and A.I.Nikishov, Yad. Fiz. {\bf 11}, 1072 (1970). \\
\item A.I.Nikishov, Nucl. Phys. {\bf B 21}, 346 (1970). \\
\item  A.Hansen, F.Ravndal, Physica Scripta, {\bf 23}, 1036 (1981).      \\
\item  W.Greiner, B.M\"uller, J.Rafelski, {\sl Quantum Electrodynamics of Strong
    Field}, Springer-Verlag (1985). \\
\item  A.Calogeracos, N.Dombey, Contemp. Phys. {\bf 40}, 313 (1999).         \\
\item A.I.Nikishov, Teor. Mat. Fiz. {\bf 98}, 60 (1994).\\
\item  {\sl Higher Transcendental Functions (Bateman Manuscript Project
     )}, Ed. by  A.Erd\'elyi (McGraw-Hill, New York, 1953; Nauka, Moscow,
    1980; Pergamon, Oxford, 1982), Vol 2. \\
\item N.B.Narozhny and A.I.Nikishov, Teor.  Mat. Fiz. {\bf 26}, 16 (1976).  \\
\item N.B.Narozhny and A.I.Nikishov, see reference 5. p. 175.                 \\
\item J.Schwinger, Phys. Rev. {\bf 82}, 664 (1951).                            \\
\item I.S.Gradstein, I.M.Ryzhik, {\sl Tables of Integrals, Sums, Series,
    and Products }, Moscow, 1962.                    \\
\item A.I.Nikishov, ZhETF, {\bf 120}, 227 (2001).\\
\item A.P.Prudnikov, Yu.A.Brychkov, and O.I.Marichev, {\sl Integrals and
    Series}, Vol I, ({\sl Elementary  Functions})
(Gordon and Breach, New York, 1986) [Russ. original, Nauka,  Moscow, 1981].
 \end{enumerate}

\end{document}